\documentclass{article}
\usepackage{graphicx}
\usepackage{amssymb}
\usepackage{amsmath}

\pdfoutput=1

\input{tcilatex}
\begin{document}

\title{Classical resonance interactions and Josephson junction in
macroscopic quantum dynamics}
\author{V.N. Pilipchuk \\
\textit{Wayne State University, Detroit, MI}\\
\textit{e-mail: pilipchuk@wayne.edu}}
\maketitle

\begin{abstract}
It is shown that the classical dynamics of 1:1 resonance interaction between
two identical linearly coupled Duffing oscillators is equivalent to the
symmetric (non-biased) case of `macroscopic' quantum dynamics of two weakly
coupled Bose-Einstein condensates. The analogy develops through the boson
Josephson junction equations, however, reduced to a single conservative
energy partition (EP) oscillator. The derived oscillator is solvable in
quadratures, furthermore it admits asymptotic solution in terms of
elementary functions after transition to the action-angle variables. Energy
partition and coherency indexes are introduced to provide a complete
characterization of the system dynamic states through the state variables of
the EP oscillator. In particular, nonlinear normal and local mode dynamics
of the original system associate with equilibrium points of such oscillator.
Additional equilibrium points - the local modes - may occur on high energy
level as a result of the symmetry breaking bifurcation, which is equivalent
to the macroscopic quantum self-trapping effect in \textit{boson} Josephson
junction. Finally, since the Hamiltonian of EP oscillator is always
quadratic with respect its linear momentum, the idea of second quantization
can be explored without usual transition to the rigid pendulum approximation.
\end{abstract}

\section{INTRODUCTION}

The present study brings attention to the analogy between 1:1 resonance
dynamics of coupled classical Duffing oscillators and macroscopic quantum
dynamics of coupled Bose-Einstein condensates \cite{Dalfovo:1999RevModPhys}.
Despite of quite different physical contents the resultant equations for
both types of interaction appear to have surprisingly the same form of boson
Josephson junction (BJJ) tunneling equations \cite{Smerzi:1997} \cite%
{Raghavan:1999}. These represent a classical single degree-of-freedom
Hamiltonian system whose conjugate states are the fractional population
disbalance and phase difference. In macroscopic quantum dynamics, such
equations are eventually obtained from the nonlinear Schr\"{o}dinger
(Gross-Pitaevskii) equation \cite{Gross:1961} \cite{Pitaevskii:1961} by
representing the corresponding wave function as a superposition of two wave
functions \cite{Smerzi:1997} \cite{Raghavan:1999}. In the case of classical
resonance dynamics, the BJJ equations are obtained after specific coordinate
transformation by applying parameter variation and conventional averaging
techniques to the differential equations of motion of interacting
oscillators. Further analogies are revealed by establishing the links
between qualitative features of the BJJ equations and the corresponding
physical effects developed in both classical and macroscopic quantum
dynamics. In particular, specifics of the energy exchange between nonlinear
oscillators due to their 1:1 resonance and so-called nonlinear normal mode
motions are discussed first. It is known that, during the normal mode
motions \cite{Scott:1985ChemPhysLett} \cite{Vakakis:1996Wiley}, the energy
partition between the oscillators from one cycle to another is fixed so that
there in no energy flow between the oscillators. However, when the initial
states of oscillators are out of compliance with any of the normal modes,
the oscillators slowly exchange by some portion or even all of the energy in
a beat-wise manner. Such beating phenomena have been in focus of nonlinear
physics and physical mechanics for few decades by very different theoretical
and practical reasons \cite{Kosevich:1989Kiev} \cite{Holm:2002SIAM}.
Usually, the beat dynamics are described through the transition to
amplitude-angle or similar coordinates followed by averaging the new
equations over one cycle of vibration. The resultant system may appear to
have one or may be several integrals dictated by the system' symmetry \cite%
{Holm:2002SIAM}. These types of integrals actually provide the description
for the energy exchange between the interacting oscillators. Recently, based
on the detailed parametric study of phase trajectories, the notion of
limiting phase trajectory was introduced \cite{Manevitch:2005} \cite%
{Manevitch:2006Samos} \cite{Manevitch:2007ArchAM}. It was noticed that, in
the limit when the entire energy of the system swings from one oscillator to
another, the descriptive phase angles resemble state variables of impact
oscillators with non-smooth temporal shapes. The importance of this
observation is that it provides asymptotic simplifications for the cases
extremely opposite to the normal modes. Further, analytical algorithms of
nonsmooth temporal transformations \cite{Pilipchuk:1996JSV} \cite%
{Pilipchuk:2010Springer} where adapted to build approximate analytical
solutions for limiting phase trajectories \cite{Manevitch:2009arXiv} \cite%
{Manevitch:2010PhysD} \cite{Starosvetsky:2011}. In the present work, it is
shown that the phase variable, which determines the energy partition (EP)
between the two oscillators, is given by a strongly nonlinear conservative
oscillator, which is exactly solvable for any intensity of the energy
exchange. Furthermore, the EP oscillator admits two different asymptotic
limits associated with the vanishing or maximal possible intensities of
beats in which the oscillator becomes harmonic or vibro-impact,
respectively. In particular, the impact limit corresponds to the limiting
phase trajectory on which the energy exchange involves the total energy of
the system. Finally, it is shown that first integral of the EP oscillator
completely determines necessary and sufficient conditions for the nonlinear
mode localization phenomenon \cite{Scott:1985ChemPhysLett} \cite%
{Vakakis:1996Wiley}. It is shown that vibrations near nonlinear local modes
of coupled oscillators are equivalent to the macroscopic quantum
self-trapping (MQST) effect. As discussed earlier \cite{Raghavan:1999},
there are two types of MQST, such as running-phase MQST and $\pi $-phase
MQST. Note that the BJJ captures both types of MQST whereas the
superconductor Josephson junction (SJJ) rigid pendulum approximation ignores
the $\pi $-phase MQST. An early overview and introduction to SJJ can be
found in \cite{Barone:1982physics}. According to this theory, a single wave
function associates with a macroscopic number of electrons which are assumed
to condense in the same quantum (superconductive) state. When two
superconductors are close enough to each other, practically at about 30 \AA %
, the superconductors begin to interact through the tunneling effect such
that at about 10 \AA\ between the superconductors their macroscopic quantum
phases cannot be viewed as independent any more and must be described as a
single system \cite{Barone:1982physics}. In a similar way, it is impossible
to separately consider two even weakly coupled classical oscillators when
they resonate. This is due to the fact that slowly oscillating resonance
energy flows between the oscillators cannot be ignored anymore and must be
interpreted as a new system. As mentioned above, such new system represents
a strongly nonlinear conservative oscillator with quite interesting physical
and mathematical properties. \ Surprisingly enough, the basic case of such
oscillator has been known for several decades as unique example of strongly
nonlinear exactly solvable oscillator with no relation to any physically
meaningful situation \cite{Kauderer:1958Springer}, except few
phenomenological applications \cite{Nesterov:1978PMIPE} \cite%
{Dimentberg:2000RSL} \cite{Pilipchuk:2008ND4}. Sections \ref{Formulation}
through \ref{localization} describe the methodology based on the classical
model of coupled Duffing oscillators. In Section \ref{BJJ} it is shown that
the resultant 1:1 classic resonance equations take the form of BJJ
equations. Then, the BJJ derivation in macroscopic quantum dynamics is
briefly recalled for comparison. reason. \ Finally, the analogies and
physical meaning of the BJJ variables in classical and quantum dynamical
cases are discussed.

\section{NONLINEAR BEATS}

\label{Formulation}

\subsection{Coupled nonlinear oscillators}

Let us consider a system of two identical linearly coupled unit-mass
oscillators (Fig.1) described by Hamiltonian

\begin{equation}
H=\frac{1}{2}\left( v_{1}^{2}+v_{2}^{2}\right) +\Pi (u_{1})+\Pi (u_{2})+%
\frac{1}{2}b\left( u_{1}-u_{2}\right) ^{2}  \label{H1}
\end{equation}
where $u_{i}$ and $v_{i}$ ($i=1,2$) are the coordinates and linear momenta,
respectively, $\Pi (u_{i})$ is an even analytic function describing the
potential energy for each of the two oscillators, and $b$ is the coupling
stiffness, which is assumed to be relatively week, $b/\Pi ^{\prime \prime
}(0)\ll 1$.

Introducing the parameters, $\Omega =\sqrt{\Pi ^{\prime \prime }(0)+b}$ and $%
\varepsilon =b\Omega ^{-2}$, and subtracting the parabolic component from
the potential energy, 
\begin{equation}
\Pi (u)-\frac{1}{2}\Pi ^{\prime \prime }(0)u^{2}\equiv \varepsilon U(u)
\label{U_def}
\end{equation}
brings Hamiltonian (\ref{H1}) to the form, which incorporates the additional
assumption that both nonlinearity and coupling are of the same order of
magnitude, $\varepsilon $, 
\begin{equation}
H=\frac{1}{2}\left( v_{1}^{2}+v_{2}^{2}\right) +\frac{1}{2}\Omega ^{2}\left(
u_{1}^{2}+u_{2}^{2}\right) +\varepsilon \left[ U(u_{1})-\Omega
^{2}u_{1}u_{2}+U(u_{2})\right]  \label{H2}
\end{equation}

Note also that the coupling is represented now in somewhat canonical form
after the non-coupled terms of the interaction energy (\ref{H1}), $%
bu_{i}^{2}/2$, have been associated with the corresponding oscillators. The
differential equations of motion are given by 
\begin{equation}
\dot{u}_{i}=\frac{\partial H}{\partial v_{i}}\text{,\quad }\dot{v}_{i}=-%
\frac{\partial H}{\partial u_{i}}  \label{eqns1}
\end{equation}
or 
\begin{eqnarray}
\dot{u}_{1} &=&v_{1}  \notag \\
\dot{u}_{2} &=&v_{2}  \notag \\
\dot{v}_{1} &=&-\Omega ^{2}u_{1}+\varepsilon \lbrack \Omega
^{2}u_{2}-U^{\prime }(u_{1})]  \label{nb3} \\
\dot{v}_{2} &=&-\Omega ^{2}u_{2}+\varepsilon \lbrack \Omega
^{2}u_{1}-U^{\prime }(u_{2})]  \notag
\end{eqnarray}

As $\varepsilon \rightarrow 0$, system (\ref{nb3}) degenerates into two
identical harmonic oscillators whose total energies are separately
conserved. At non-zero $\varepsilon $, the oscillators become non-linear and
coupled with each other in such way that one of the oscillators is loaded
proportionally to the displacement of another oscillator. Since system (\ref%
{nb3}) is perfectly symmetric and conservative, it is reasonable to assume a
relatively slow energy exchange between the oscillators due to the weak
coupling. In order to describe the energy exchange dynamics in physically
meaningful terms, let us introduce a new set of variables as follows $%
\{u_{1},v_{1},u_{2},v_{2}\}->\{K(t),\theta (t),\delta (t),\Delta (t)\}$: 
\begin{eqnarray}
u_{1} &=&\sqrt{K}\cos \left( \frac{\theta }{2}+\frac{\pi }{4}\right) \cos
\delta  \notag \\
v_{1} &=&-\sqrt{K}\Omega \cos \left( \frac{\theta }{2}+\frac{\pi }{4}\right)
\sin \delta  \notag \\
u_{2} &=&-\sqrt{K}\sin \left( \frac{\theta }{2}+\frac{\pi }{4}\right) \cos
(\delta +\Delta )  \label{nb4} \\
v_{2} &=&\sqrt{K}\Omega \sin \left( \frac{\theta }{2}+\frac{\pi }{4}\right)
\sin (\delta +\Delta )  \notag
\end{eqnarray}

If $K$, $\theta $, and $\Delta $ are constant, and $\delta =\Omega t$,
expressions (\ref{nb4}) represent the exact general solution of the
decoupled set of harmonic oscillators (\ref{nb3}), $\varepsilon =0$.
Therefore, relationships (\ref{nb4}) implement the idea of parameter
variations compensating the perturbation, when $\varepsilon \neq 0$. In
order to track the oscillator energies during the vibration process, let us
introduce quantities 
\begin{eqnarray}
E_{1} &=&\frac{1}{2}(v_{1}^{2}+\Omega ^{2}u_{1}^{2})=\frac{1}{2}E_{0}(1-\sin
\theta )  \notag \\
E_{2} &=&\frac{1}{2}(v_{2}^{2}+\Omega ^{2}u_{2}^{2})=\frac{1}{2}E_{0}(1+\sin
\theta )  \label{nb5} \\
E_{12} &=&\frac{1}{2}(v_{1}v_{2}+\Omega ^{2}u_{1}u_{2})=-\sqrt{E_{1}E_{2}}%
\cos \Delta  \notag
\end{eqnarray}
where 
\begin{equation}
E_{0}=E_{1}+E_{2}=\frac{1}{2}\Omega ^{2}K  \label{nb6}
\end{equation}

Expressions (\ref{nb5}) and (\ref{nb6}) clarify physical meaning of the
variables $K$, $\theta $ and $\Delta $ participating in transformation (\ref%
{nb4}), while the variable $\delta $ is the fast phase associated with the
principal temporal rate of the vibrating oscillators. In particular, the
quantity $K$ is proportional to the total energy of the decoupled and
linearized oscillators, the phase $\theta $ characterizes the energy
partition between the oscillators, and $\Delta $ relates to the phase shift
between the oscillators as discussed below. In case $\varepsilon \neq 0$,
the energy parameter $K$ has small temporal fluctuations due to the coupling
and nonlinear terms in (\ref{nb3}). Nevertheless, expressions (\ref{nb5})
and (\ref{nb6}) can still be used for characterization of the energy
distribution between the oscillators. For that purpose, as follows from (\ref%
{nb5}), the interval $-\pi /2\leq \theta \leq \pi /2$, on which $E_{0}\geq
E_{1}\geq 0$ and $0\leq E_{2}\leq E_{0}$, is sufficient. In particular, the
case $\theta =0$ corresponds to equipartition, $E_{1}=E_{2}$, under which
the system oscillate either out--of-phase ($\Delta =0$) or in-phase ($\Delta
=\pi $) according to the sign convention in (\ref{nb4}). It is convenient to
deal with one full period of the phase shift $\Delta $ within the range $%
-\pi /2\leq \Delta \leq 3\pi /2$, including both in-phase and out-of-phase
modes.

\subsection{Energy partition and coherency indexes}

As an alternative to the angular quantities $\theta $ and $\Delta $, let us
introduce the \textit{energy partition index} $P$ and \textit{coherency index%
} $Q$ describing the system' vibrating states as follows 
\begin{equation}
P=\frac{E_{1}-E_{2}}{E_{1}+E_{2}}=-\sin \theta \text{,\quad }-1\leq P\leq 1
\label{partition_def}
\end{equation}
\begin{equation}
Q=\frac{E_{12}}{\sqrt{E_{1}E_{2}}}=-\cos \Delta \text{,\quad }-1\leq Q\leq 1
\label{coherency_def}
\end{equation}

According to definition (\ref{partition_def}), the number $P=0$ indicates
the energy equipartition, $E_{1}=E_{2}$, whereas $P=1$ or $P=-1$ correspond
to the case when all the energy belongs to either first or second
oscillator, respectively. According to definition (\ref{coherency_def}), the
boundaries $Q=1$ and $Q=-1$ correspond to in-phase and out-of-phase modes,
respectively, as illustrated by Figs. 1 and 2.

The derivation of equations for the descriptive phase variables includes
substitution of the coordinate transformation (\ref{nb4}) in (\ref{nb3}),
which is in fact the variation of constants procedure, and then implementing
the averaging with respect to the fast phase $\delta $ as described in
Appendix. Assuming the monomial form, $U(u_{i})=\alpha u_{i}^{4}/4$, such
averaging gives 
\begin{eqnarray}
\dot{K} &=&0  \notag \\
\dot{\theta} &=&\varepsilon \Omega \sin \Delta  \notag \\
\dot{\Delta} &=&-\varepsilon \Omega (\cos \Delta \tan \theta -\kappa \sin
\theta )  \label{nb8} \\
\dot{\delta} &=&\Omega +\frac{1}{2}\varepsilon \Omega \left[ \cos \Delta
\tan \left( \frac{\theta }{2}+\frac{\pi }{4}\right) +\kappa (1-\sin \theta )%
\right]  \notag
\end{eqnarray}
where the following nonlinearity parameter is introduced 
\begin{equation}
\kappa =\frac{3\alpha K}{8\Omega ^{2}}  \label{kappa}
\end{equation}

The first equation in (\ref{nb8}) shows that the energy parameter $K$
remains averagely constant regardless the perturbation parameter $%
\varepsilon $. The fact that $K$ is constant justifies the use of quantities
(\ref{nb5}) and (\ref{nb6}) for characterization of the energy exchange
between the oscillators since neither the coupling nor nonlinear stiffness
in (\ref{nb3}) can accumulate the energy during one vibration cycle. In
order to clarify physical meaning of the parameter $\kappa $, consider a
single oscillator, $\ddot{u}+\Omega ^{2}u+\alpha u^{3}=0$, whose mean (over
the period) potential energy components, corresponding to linear and
nonlinear stiffness terms, are $E_{\Omega }=\Omega ^{2}<u^{2}>/2$ and $%
E_{\alpha }=\alpha <u^{4}>/4$, respectively. Assuming the harmonic temporal
mode for the coordinate $u(t)$, and taking into account (\ref{nb6}) and (\ref%
{kappa}), gives 
\begin{equation}
\kappa =\frac{E_{\alpha }}{E_{\Omega }}  \label{kappa1}
\end{equation}

Therefore, $\kappa $ characterizes the strength of nonlinearity in terms of
its relative energy capacity during one vibration cycle. Taking into account
(\ref{nb8}) and (\ref{kappa}), gives also $\dot{\kappa}=0$. As follows from (%
\ref{nb8}), further complete description of the dynamics can be conducted
now in terms of the two phase shift parameters, $\Delta (t)$ and $\theta (t)$%
; the fast phase $\delta (t)$ is obtained then by integration from the last
equation in (\ref{nb8}).

As discussed above, the corresponding dynamical system on the phase plane $%
\Delta -\theta $ will be considered on the rectangular 
\begin{equation}
D=\{-\pi /2\leq \Delta \leq 3\pi /2,-\pi /2\leq \theta \leq \pi /2\}
\label{D}
\end{equation}

In the low nonlinearity range, $0\leq \kappa <1$, there are two stationary
points, $(0,0)$ and $(\pi ,0)$, corresponding to the out-of-phase and
in-phase vibration modes, respectively. However, two more stationary points $%
(\pm \arccos \kappa ^{-1},0)$ emerge from zero when the nonlinearity becomes
sufficiently strong, $1\leq \kappa $; see Fig. 1. Geometrical interpretation
of the bifurcation at $\kappa =1$ will be given below in Section \ref{Joseph}%
.

\section{EP OSCILLATOR}

\label{EPOscillator}

It can be shown by inspection that system (\ref{nb8}) admits the integral as
follows 
\begin{equation}
G\equiv -\cos \Delta \cos \theta +\frac{1}{2}\kappa \cos ^{2}\theta =const.
\label{nb11}
\end{equation}

Taking into account (\ref{nb11}) and eliminating the phase $\Delta $ from
the second and third equations of system (\ref{nb8}), gives a single
strongly nonlinear conservative oscillator with respect to the coordinate $%
\theta $ in the form \cite{Pilipchuk:2008ND4}, \cite{Pilipchuk:2010Springer} 
\begin{equation}
\ddot{\theta}+(\varepsilon \Omega )^{2}\left( G^{2}\frac{\tan \theta }{\cos
^{2}\theta }-\frac{1}{8}\kappa ^{2}\sin 2\theta \right) =0  \label{nb13}
\end{equation}

Note that the quantity $G$ in equation (\ref{nb13}) remains constant only on
a fixed dynamic trajectory in the plane $\theta $-$\Delta $ but may vary
from one trajectory to another. Therefore, the number $G$ (\ref{nb11}) must
be calculated first by fixing some point $\{\theta _{0}$,$\Delta _{0}\}$ on
the trajectory. Then equation (\ref{nb13}) can be solved by making sure that
the initial condition $\{\theta (0)$, $\dot{\theta}(0)\}$ corresponds to the
fixed trajectory according to the second equation in (\ref{nb8}). The
parameter $\kappa $ however can be chosen regardless the dynamics in the
plane $\theta $-$\Delta $. According to (\ref{nb5}), equation (\ref{nb13})
constitutes a principal equation describing the energy exchange between
oscillators (\ref{nb3}). Moreover, if the function $\theta (t)$ is known,
then other two phase variables, $\Delta $ and $\delta $, are obtained from
system (\ref{nb8}) by differentiation and integration.

It is shown therefore that the entire system (\ref{nb8}) is exactly solvable
in quadratures since general solution of equation (\ref{nb13}) can be
obtained from its `energy' integral $H_{\theta }=const$, 
\begin{equation}
H_{\theta }\equiv \frac{1}{2}\dot{\theta}^{2}+(\varepsilon \Omega
)^{2}\left( \frac{1}{2}G^{2}\tan ^{2}\theta +\frac{1}{16}\kappa ^{2}\cos
2\theta \right)  \label{Htheta}
\end{equation}

The dynamics of oscillator (\ref{nb13}) essentially depends on the shape of
its potential energy within the interval $-\pi /2<\theta <\pi /2$. In
particular, if the parameter $\kappa $ is small enough, then oscillator (\ref%
{Htheta}) has one stable equilibrium position at $\theta =0$. However, high
energy levels of the original system (\ref{H1}), that is large $\kappa $,
can make the equilibrium position $\theta =0$ unstable by generating two new
(stable) equilibrium positions. Such kind of bifurcation obviously relates
to that described at the end of Section \ref{Formulation}. According to
definition (\ref{partition_def}), every equilibrium of the oscillator (\ref%
{nb13}) corresponds to a fixed (no beats) energy partition between the
oscillators in the original system, in other words, - nonlinear normal mode.
Recall that oscillator (\ref{nb13}) is obtained after the procedure of
averaging has been applied to the original model (\ref{nb3}). \ For
validation purposes, Fig. 3 illustrates the behavior of energy partition
index (\ref{partition_def}) based on numerical solutions of both types of
models the original model (\ref{nb3}) and oscillator (\ref{nb13}). The
initial configurations are chosen to cover the areas of normal and local
modes as illustrated by the model trajectories on configuration planes in
Fig. 4. Overall, Figs. 4 through 7 illustrate the dynamic trajectories on
configuration planes and the corresponding behaviors of energy partition (%
\ref{partition_def}) and coherency (\ref{coherency_def}) indexes, when
increasing the energy (nonlinearity) level above the critical point $\kappa
=\kappa ^{\ast }=1$. Such diagrams give quite complete characterization of
the dynamics by showing how the energy is distributed between the
oscillators and what is the phase shift between the oscillators at any given
time. Note that the derived oscillator (\ref{nb13}) captures both local
dynamics near separate nonlinear normal modes and drifts over multiple
modes, including stable and unstable ones. This essentially complements the
theory of nonlinear normal modes in classical dynamics covering mainly
periodic normal mode motions and their small neighborhoods \cite%
{Vakakis:1996Wiley}.

\section{ASYMPTOTICS OF\ BEATS}

\subsection{Linearized case}

\label{linearcase}

When the original system is linear ($\alpha =0$ $\Longrightarrow $ $\kappa
=0 $), equation (\ref{nb13}) admits explicit analytical solution within the
class of elementary functions 
\begin{equation}
\theta (t)=\arcsin \left[ \sin \theta _{0}\sin \phi (t)\right] \text{ }
\label{nb15}
\end{equation}%
where $\theta _{0}$ is the amplitude of $\theta $, whereas another constant
can be introduced into the phase $\phi (t)=\varepsilon \sec \theta
_{0}|G|\Omega t$ as an arbitrary temporal shift admitted by equation (\ref%
{nb13}). As mentioned in Introduction, such type of explicit solution has
been known for quite a long time with no relation to any physical system 
\cite{Kauderer:1958Springer}, however it was used recently in some physical
and mechanical applications \cite{Nesterov:1978PMIPE}, \cite%
{Dimentberg:2000RSL} in a phenomenological way. Although solution (\ref{nb15}%
) holds only for the linear model, $\kappa =0$ , it nevertheless helps to
clarify specifics of the behavior of phase variables in nonlinear cases. In
particular, substituting (\ref{nb15}) in the second equation of (\ref{nb8}),
gives 
\begin{equation}
\Delta (t)=\arcsin \left[ \frac{|G|\tan \theta _{0}\cos \phi (t)}{\sqrt{%
1-\sin ^{2}\theta _{0}\sin ^{2}\phi (t)}}\right]  \label{Delta}
\end{equation}

Fig. 8 illustrates the relationship between the beat dynamics of the
linearized system and the corresponding phase variables, $\theta (t)$ and $%
\Delta (t)$. The coordinates $u_{i}(t)$, ($i=1,2$) represent exact
analytical solution under the initial conditions obtained from (\ref{nb4})
at $t=0$. The integral $G$ is calculated at the amplitude point\footnote{%
The notation $\theta _{0}$ should not be confused with the initial value,
which is $\theta (0)=0$, according to the present form of the solution.}, $%
\theta _{0}=\pi /2-0.01$, chosen slightly below its maximal possible
magnitude $\pi /2$. Since $\Delta =0$ as $\theta =\theta _{0}$, then $%
G=-\cos \theta _{0}$ and therefore $\phi (t)=\varepsilon \Omega t$. Other
parameters are taken as $\varepsilon =0.01$, $\Omega =1.0$, $K=1.0$ and $%
\delta (0)=0$. As follows from Fig. 8, the behavior of phase variables $%
\theta $ and $\Delta $ resembles smoothed time histories of the coordinate
and velocity of a simple impact oscillator. As mentioned in Introduction,
this fact was noticed first in \cite{Manevitch:2007ArchAM}, \cite%
{Manevitch:2006Samos} based on the analysis of phase equations similar to (%
\ref{nb8}) however obtained in a different way after complexification of the
original coordinates. In particular, it was found that the `impact limit'
corresponds to the most intensive energy exchange between the oscillators
when each of the oscillators periodically hosts the total energy of the
system. It is seen now that, in the linearized case, such asymptotic follows
directly from exact solutions, (\ref{nb15}) and (\ref{Delta}), 
\begin{eqnarray}
\theta (t) &\rightarrow &\arcsin (\sin \phi )=\frac{\pi }{2}\tau \left( 
\frac{2}{\pi }\phi \right)  \notag \\
\Delta (t) &\rightarrow &\arcsin \left( \frac{\cos \phi }{|\cos \phi |}%
\right) =\frac{\pi }{2}e\left( \frac{2\phi }{\pi }\right)  \label{impact1} \\
\phi &=&\varepsilon \Omega t\text{,~ }\theta _{0}\rightarrow \pi /2  \notag
\end{eqnarray}%
where $\tau (z)$ and $e(z)$ are triangular sine and rectangular cosine wave
functions whose amplitude is unity and the period is normalized to four in
order to provide the basic relationships of nonsmooth temporal
transformations \cite{Pilipchuk:2010Springer}, $\tau ^{\prime }(z)=e(z)$ and 
$e^{2}(z)=1$.

Now substituting (\ref{nb15}) in (\ref{nb5}), gives

\begin{eqnarray}
E_{1} &=&\frac{1}{2}E_{0}(1-\sin \theta _{0}\sin \varepsilon \Omega t) 
\notag \\
E_{2} &=&\frac{1}{2}E_{0}(1+\sin \theta _{0}\sin \varepsilon \Omega t)
\label{impact2}
\end{eqnarray}

Taking into account (\ref{partition_def}) and (\ref{impact2}), gives the
corresponding energy partition index 
\begin{equation}
P(t)=-\sin \theta _{0}\sin \varepsilon \Omega t  \label{partition}
\end{equation}

Recall that the number $P=0$ indicates equipartition, $E_{1}=E_{2}$, whereas 
$P=1$ or $P=-1$ correspond to the case when all the energy belongs to either
first or second oscillator, respectively. As follows from (\ref{partition}),
such states can be reached only in the limit case (\ref{impact1}), when $%
\theta _{0}=\pi /2$ .

In case of small amplitudes $|\theta _{0}|\ll \pi /2$, corresponding to a
moderate energy exchange, solutions (\ref{nb15}) and (\ref{Delta}) are
approaching another simple limit of harmonic temporal shapes as illustrated
by Fig. 9, where the amplitude is $\theta _{0}=0.5$. This also follows
directly from the linearization of equation (\ref{nb13}) near zero $\theta
=0 $.

\subsection{Nonlinear case}

\label{actionangle}

In this subsection, explicit analytical solution of the EP oscillator is
obtained via transition to the action-angle variables, $\{\theta $,$\theta
^{\prime }\}\longrightarrow \{I$,$\phi \}$, of the generating model, $\kappa
=0$. Re-scaling the time, $p=\varepsilon \Omega Gt$, brings the effective
Hamiltonian of EP oscillator (\ref{Htheta}) to the form 
\begin{equation}
H_{p}=\frac{H_{\theta }}{(\varepsilon \Omega G)^{2}}=\frac{1}{2}(\theta
^{\prime 2}+\tan ^{2}\theta +\mu \cos 2\theta )  \label{Hp}
\end{equation}%
where $\theta ^{\prime }=d\theta /dp$ is interpreted as a linear momentum, $%
\mu =\kappa ^{2}/(8G^{2})$ is another parameter associated with the
nonlinearity of Duffing oscillators $\kappa $ defined by (\ref{kappa}), and
the quantity $G$ given by (\ref{nb11}) must be treated as a fixed number. \ 

Based on the exact solution (\ref{nb15}), the action-angle variables are
introduced as follows \cite{Pilipchuk:2008ND4}, \cite{Pilipchuk:2010Springer}
\begin{eqnarray}
\theta &=&\arcsin \left( \frac{\sqrt{2I+I^{2}}}{1+I}\sin \phi \right)  \notag
\\
\theta ^{\prime } &=&\frac{\left( 1+I\right) \sqrt{2I+I^{2}}\cos \phi }{%
\sqrt{1+\left( 2I+I^{2}\right) \cos ^{2}\phi }}  \label{nb25}
\end{eqnarray}

Now canonical transformation (\ref{nb25}) brings Hamiltonian (\ref{Hp}) and
the corresponding differential equations of motion to the form 
\begin{equation}
H_{p}=I+\frac{1}{2}I^{2}+\frac{\mu }{2}\frac{1+I(2+I)\cos 2\phi }{(1+I)^{2}}
\label{Hp1}
\end{equation}

and 
\begin{eqnarray}
\frac{dI}{dp} &=&-\frac{\partial H_{p}}{\partial \phi }\equiv \mu \frac{%
I(2+I)}{(1+I)^{2}}\sin 2\phi  \notag \\
\frac{d\phi }{dp} &=&\frac{\partial H_{p}}{\partial I}\equiv 1+I-\frac{2\mu 
}{(1+I)^{3}}\sin ^{2}\phi  \label{nb26}
\end{eqnarray}

The essential advantage of new system (\ref{nb26}) is that it becomes linear
as $\mu =0$ while the original oscillator (\ref{Hp}) still remains strongly
nonlinear. The idea of averaging can be also implemented as asymptotic
integration of system (\ref{nb26}) by means of the coordinate transformation 
$\{I,\phi \}\longrightarrow \{J,\psi \}$: 
\begin{eqnarray}
I &=&J-\mu \frac{J(2+J)}{2(1+J)^{3}}\cos 2\psi +O(\mu ^{2})  \notag \\
\phi &=&\psi -\mu \frac{(J^{2}+2J-2)}{4(1+J)^{4}}\sin 2\psi +O(\mu ^{2})
\label{nb27}
\end{eqnarray}

Transformation (\ref{nb27}) is obtained from the condition eliminating the
fast phase $\varphi $ from the terms of order $\mu $ on the right-hand side
in such a way that the new system takes the form 
\begin{eqnarray}
\frac{dJ}{dp} &=&O(\mu ^{2})  \notag \\
\frac{d\psi }{dp} &=&1+J-\frac{\mu }{(1+J)^{3}}+O(\mu ^{2})  \label{nb28}
\end{eqnarray}
System (\ref{nb28}) is easily integrated as follows 
\begin{eqnarray}
J &=&J_{0}  \label{nb29} \\
\psi &=&\left[ 1+J-\frac{\mu }{(1+J)^{3}}\right] p+\psi _{0}  \notag
\end{eqnarray}
where $J_{0}$ and $\psi _{0}$ are constants of integration.

Finally, substituting (\ref{nb29}) in (\ref{nb27}) and then (\ref{nb27}) in (%
\ref{nb25}), gives solution $\theta (p)$ whose effectiveness is illustrated
by Figs. 10 and 11 at different nonlinearity and intensity of energy
exchange levels, $\kappa $ and $P_{0}$, respectively. In particular, the
transition from harmonic to sawtooth temporal shape of the phase $\theta $
is due to the increase of the amplitude of energy partition index, $P_{0}$.
The nonlinearity parameter $\kappa $ has certain effect on the curvature of
lines as seen from Fig. 11. Note that, although the above procedure of
asymptotic integration assumes the parameter $\mu (\kappa )$ to be small as
compared to unity, the solution appears to be very effective also under
quite significant magnitudes of $\mu $ provided that the corresponding
trajectory holds its symmetry on the phase plane $\theta $-$\theta ^{\prime
} $. However, the error of solution is growing as its trajectory becomes
close to the separatrix of EP oscillator, which occurs under the condition $%
\kappa >1$; see Section \ref{localization} for physical interpretation. Such
situation appears to be quite common due to highly sensitive dynamics near
separatrix loop.

Besides, expressions (\ref{nb27}) and (\ref{nb29}) point to the fact that,
as the parameter $\mu $ increases, the monotonic growth of angle coordinate $%
\phi $ can be broken. This is confirmed also by the direct numerical
integration of equations (\ref{nb26}) at different magnitudes of the
parameter $\mu $ as illustrated by Fig. 12. The initial angle is $\phi =\pi
/2$ that associates with the extremum of the energy partition index $P$. \
The new action variable is fixed as $J=1.0$, while the parameter $\mu $ is
incrementally increased. As follows from the notations, this is eventually
equivalent to incremental change of both the energy parameter $\kappa $ and
the amplitude of energy partition index, $P_{0}$. Detailed calculations show
that above approximately $\mu =3.559$ ($\kappa =1.4549$ and $P_{0}=-0.927062$%
) the angle coordinate locks near its initial value $\phi =\pi /2$, as a
result trajectories on the planes represented by Fig. 12 become closed.

\section{MODE LOCALIZATION}

\label{localization}

The mode localization effect is shown schematically in Fig. 1. As mentioned
in Section \ref{EPOscillator}, the equilibrium points of oscillator (\ref%
{nb13}) represent nonlinear normal modes of the original system. Although,
in the case $\kappa \neq 0$, oscillator (\ref{nb13}) is still solvable in
quadratures, let us consider the following qubic approximation, by assuming
that $|\theta |\ll \pi /2$, 
\begin{equation}
\ddot{\theta}+(\varepsilon \Omega )^{2}\left[ \left( G^{2}-\frac{\kappa ^{2}%
}{4}\right) \theta +\frac{1}{6}\left( 8G^{2}+\kappa ^{2}\right) \theta ^{3}%
\right] =0  \label{Duffing}
\end{equation}

As follows from (\ref{nb5}), the equilibrium point $\theta =0$ of oscillator
(\ref{Duffing}) corresponds to the equal energy distribution, under which
the original model (\ref{H1}) remains in one of its two symmetric nonlinear
normal modes. So when the linear stiffness is positive, equation (\ref%
{Duffing}) has periodic solutions describing the energy exchange between
oscillators (\ref{nb3}) near in-phase or out-of-phase mode. However, the
equal energy distribution, associated with the equilibrium $\theta =0$,
becomes unstable if the linear stiffness is negative, $G^{2}-\kappa ^{2}/4<0$%
. In this case two new stable equilibria surrounded by separatrix loops
occur near the unstable equilibrium. This indicates the onset of nonlinear
local modes of the original system (\ref{nb3}) with a sustainable disbalance
in the energy distribution despite of the perfect symmetry of system (\ref%
{H1}). In terms of the present notations, the condition of negative linear
stiffness can be represented in the form \cite{Pilipchuk:2010Springer} 
\begin{equation}
f_{2}\equiv -\frac{\kappa (2-P_{0}^{2})}{2\sqrt{1-P_{0}^{2}}}<Q_{0}<\frac{%
\kappa P_{0}^{2}}{2\sqrt{1-P_{0}^{2}}}\equiv f_{1}  \label{necessary}
\end{equation}%
where $P_{0}=-\sin \theta _{0}$ and $Q_{0}=-\cos \Delta _{0}$ the initial
energy partition index as defined by (\ref{partition_def}).

Condition (\ref{necessary}) constitutes a necessary condition of
localization because it does not guarantee that the dynamics is trapped
inside one of the separatrix loops. The corresponding sufficient condition
is obtained from the energy integral of oscillator (\ref{Duffing}) in the
form inequalities, first of which never holds for positive $\kappa $, 
\begin{eqnarray}
Q_{0} &>&\frac{2+\kappa P_{0}^{2}}{2\sqrt{1-P_{0}^{2}}}\equiv g_{1}
\label{sufficient} \\
Q_{0} &<&-\frac{2-\kappa P_{0}^{2}}{2\sqrt{1-P_{0}^{2}}}\equiv g_{2}  \notag
\end{eqnarray}

Both estimates (\ref{necessary}) and (\ref{sufficient}) are also valid
locally, in the neighborhood of zero $\theta =0$, for strongly nonlinear
oscillator (\ref{Htheta}).

The above conditions (\ref{necessary}) and (\ref{sufficient}) must be
considered under the obvious constraint $|Q_{0}|\leq 1$. Fig. 13 illustrates
a relatively low nonlinearity case, when localization is impossible. The
solid lines represent the boundary functions, introduced in (\ref{necessary}%
) and (\ref{sufficient}), whereas the couple of dashed lines indicates the
rectangular area within which solutions of inequalities (\ref{necessary})
and (\ref{sufficient}) the above mentioned constraint. When the strength of
nonlinearity $\kappa $ is increased, the line $g_{2}$ moves upward, whereas
the line $f_{2}$ moves downward. When passing one through another at about $%
\kappa =1$, two small areas of localization occur as shown in Fig. 14. Note
that, in both localization areas, the initial phase angle $\Delta $ lays in
the neighborhood of zero. Therefore, according to (\ref{nb4}), the localized
modes branch out of the out-of-phase modes as the nonlinearity becomes
sufficiently strong.

\section{THE\ BJJ FORMALISM}

\label{BJJ}

\subsection{Classical resonance of coupled oscillators}

\label{Joseph}

The advantage of using the phase angle $\theta $ develops through the form
of oscillator (\ref{nb13}), which, in particular, admits clear qualitative
analyses and allows for explicit analytical solution in terms of elementary
functions when $\kappa =0$. This fact is essentially employed in Section \ref%
{actionangle}. In this section, however, the energy partition index $P$ will
be used instead of the phase angle $\theta $ in order to track the analogies
with conventional approaches of macroscopic quantum dynamics as discussed in
Section \ref{quantumform} below. Taking into account (\ref{partition_def})
and eliminating the angle $\theta $ from (\ref{nb4}) and (\ref{nb8}), gives
the coordinate transformation and the resultant differential equations in
the form, respectively, 
\begin{eqnarray}
u_{1} &=&\sqrt{\frac{1}{2}K(1+P)}\cos \delta  \notag \\
v_{1} &=&-\Omega \sqrt{\frac{1}{2}K(1+P)}\sin \delta  \notag \\
u_{2} &=&-\sqrt{\frac{1}{2}K(1-P)}\cos (\delta +\Delta )  \label{nb4_P} \\
v_{2} &=&\Omega \sqrt{\frac{1}{2}K(1-P)}\sin (\delta +\Delta )  \notag
\end{eqnarray}%
and 
\begin{eqnarray}
\dot{P} &=&-\varepsilon \Omega \sqrt{1-P^{2}}\sin \Delta \equiv -\frac{%
\partial H_{eff}}{\partial \Delta }  \label{nb8P} \\
\dot{\Delta} &=&-\varepsilon \Omega \left( \kappa P-\frac{P\cos \Delta }{%
\sqrt{1-P^{2}}}\right) \equiv \frac{\partial H_{eff}}{\partial P}  \notag
\end{eqnarray}%
where 
\begin{eqnarray}
H_{eff} &=&-\varepsilon \Omega \left( \frac{1}{2}\kappa P^{2}+\cos \Delta 
\sqrt{1-P^{2}}\right)  \label{Heff} \\
&=&\varepsilon \Omega \left( G-\frac{1}{2}\kappa \right)  \notag
\end{eqnarray}%
and $G$ is the result of substitution (\ref{partition_def}) in (\ref{nb11}).

System (\ref{nb8P}) appears to have the form of BJJ describing the
interaction of two Bose-Einstein condensates in macroscopic quantum dynamics 
\cite{Smerzi:1997} \cite{Raghavan:1999}. Further details of this analogy are
discussed below in the present section. Equations (\ref{nb8P}) are solved
independently on the fast phase $\delta $, which is determined then by
integration from

\begin{equation}
\dot{\delta}=\Omega +\frac{1}{2}\varepsilon \Omega \left[ \kappa (1+P)+\cos
\Delta \sqrt{\frac{1-P}{1+P}}\right]  \label{nb8Pdelta}
\end{equation}

Note that the effective Hamiltonian (\ref{Heff}) can be obtained directly by
substituting (\ref{nb4_P}) in (\ref{H2}) and averaging the result with
respect to the phase $\delta $ as follows 
\begin{equation}
H_{eff}=-\frac{4}{K\Omega }\left\langle H\right\rangle _{\delta }+2\Omega +%
\frac{1}{2}\varepsilon \Omega \kappa  \label{Heff_Have}
\end{equation}

Comparing the left-hand sides of equations (\ref{Heff}) and (\ref{Heff_Have}%
) reveals the nature of integral (\ref{nb11}), which is the constant mean
value of the Hamiltonian, $\left\langle H\right\rangle _{\delta }=const$. It
was already mentioned that, from the mathematical standpoint, the
Hamiltonian system (\ref{nb8P}) and (\ref{Heff}) with its simplifications is
similar to that usually appears in macroscopic quantum dynamics as Josephson
junction equations; see Subsection \ref{quantumform} for references and
details. It known that such systems are exactly solvable in terms of
elliptic functions \cite{Smerzi:1997}, \cite{Raghavan:1999}. Eliminating the
angle $\Delta $ in (\ref{Heff}) by means of the first equation system (\ref%
{nb8P}), gives 
\begin{equation}
\dot{P}^{2}=(\varepsilon \Omega )^{2}\left[ 1-P^{2}-\left( \frac{1}{2}\kappa
P^{2}-H_{0}\right) ^{2}\right]  \label{integral}
\end{equation}
where $H_{0}=-H_{eff}/(\varepsilon \Omega )|_{t=0}$.

Further manipulations leading to the elliptic functions are described in 
\cite{Raghavan:1999}. In particular, equation (\ref{integral}) gives the
following quadrature 
\begin{equation}
\frac{\varepsilon \Omega \kappa t}{2}=\int_{P(t)}^{P(0)}\frac{dz}{\sqrt{%
(a^{2}+z^{2})(c^{2}-z^{2})}}  \label{quadrature}
\end{equation}
where the parameters $a$ and $c$ are given by 
\begin{eqnarray*}
a^{2}c^{2} &=&\frac{4}{\kappa ^{2}}\left( 1-H_{0}^{2}\right) \\
a^{2}-c^{2} &=&\frac{4}{\kappa ^{2}}\left( 1-\kappa H_{0}\right)
\end{eqnarray*}

Below, a series of diagrams (Figs. 15 and 16) provide qualitative
descriptions of the dynamics in both $\Delta -P$ and $Q-P$ planes.

The evolution of topological structure of Hamiltonian (\ref{Heff}) with
gradually increasing nonlinearity levels is illustrated in Fig. 15. Similar
diagrams were obtained in \cite{Raghavan:1999} for the case of two weakly
coupled Bose-Einstein condensates. Classical nonlinear oscillatory chains
were analyzed in somewhat different coordinates in \cite%
{ManevitchSmirnov:2010}, and when investigating the intensity of energy
exchange between parts of periodic nonlinear Frenkel--Kontorova and
Klein--Gordon lattices based on the concept of limiting phase trajectories 
\cite{SmirnovManevitch:2011AcPh}.

As mentioned at the end of Section \ref{Formulation}, the number of
stationary points within the area, corresponding to rectangular (\ref{D}),
is increased by two when the nonlinearity parameter goes above the
bifurcation level $\kappa =1$. The new points, whose images are denoted by 
\textit{L} in Fig. 15(c), correspond to the so-called \textit{local modes}
of the coupled Duffing oscillators branching out of the out-of-phase mode 
\textit{O}; see also Fig. 1. As follows from the diagrams Fig. 15 (c-d),
near the points \textit{L}, the energy partition index $P$ preserves its
signature, in other words, the energy is accumulated in mostly one of the
two oscillators despite of the perfect symmetry of the mechanical model. If
the initial state ($\Delta (0)$,$P(0)$) lays outside but still close to the
separatrix loop, which goes through the the image of out-of-phase mode 
\textit{O}, the dynamics is not localized any more but combines the effects
of two stable local (\textit{L}) and one unstable out-of-phase (\textit{O})
modes. Such a combination becomes impossible however as further increase of
the nonlinearity level $\kappa $ leads to the structural transition in the
phase portrait such that the separatrix loop closes around the image of
in-phase mode (\textit{I}) rather than local modes (\textit{L}). As a
result, possible dynamics develop either near the in-phase mode or near one
of the two local modes. Note that at high nonlinearity levels, $\kappa >>1$,
the contour lines of effective Hamiltonian (\ref{Heff}) resemble the phase
portrait of classic pendulum as follows, for instance, from Fig. 15 (d).
Based on such similarity, it is possible to simplify the effective
Hamiltonian (\ref{Heff}) as follows 
\begin{equation}
H_{eff}\longrightarrow H_{J}=-\varepsilon \Omega \left( \frac{1}{2}\kappa
P^{2}+\cos \Delta \right)  \label{Heff_pend}
\end{equation}%
\textit{\ }

The effect of such simplification on the contour lines, however, would lead
to disappearance of stationary points \textit{L} corresponding to the local
modes. Hamiltonian (\ref{Heff_pend}) gives the equation of pendulum whose
angle is measured from the inverted (unstable) equilibrium 
\begin{equation}
\ddot{\Delta}=(\varepsilon \Omega )^{2}\kappa \sin \Delta  \label{Eqn_pend}
\end{equation}

Note that the conventional phase difference used in the macroscopic quantum
dynamics is $\Phi =\pi -\Delta $. The rigid pendulum analogy represents a
typical reduction of SJJ equations describing the interaction of two weakly
coupled superconductors, where $\Phi $ is a relative phase, and $P$ is a
fractional population imbalance \cite{Raghavan:1999}. However, the BJJ in a
double-well trap still requires the complete form of equations.

Finally, using definitions (\ref{partition_def}) and (\ref{coherency_def}),
brings the diagrams of Fig. 15 to the form represented by Fig. 16. This
gives a direct interpretation of the dynamics in terms of the energy
partition and coherency indexes. In particular, the diagrams of Fig. 16
fully comply with the understanding of nonlinear normal modes as coherent
motions of the system particles \cite{Vakakis:1996Wiley}. Namely, the normal
modes correspond to the stationary points located on the vertical lines $%
Q=\pm 1$. In other words, when the energy partition index $P$ is fixed, the
oscillators coherently vibrate either in-phase ($Q=+1$) or out-of-phase ($%
Q=-1$), according to the definition of normal modes. On the configuration
plane $u_{1}u_{2}$, such type of motions is represented by pieces of lines
passed by the system twice per one period. However, when the initial
combination $P(0)$ and $Q(0)$ does not coincide with a stationary point
then, during the vibrating process, both indexes slowly move along some
trajectory on the plane $PQ$ as shown in Fig. 16.

\subsection{Macroscopic quantum dynamics}

\label{quantumform}

For comparison reason, let us reproduce few basic steps of the BJJ
derivation in macroscopic quantum dynamics. Consider a two-level quantum
system whose dynamics is described by the discrete version of nonlinear Schr%
\"{o}dinger (Gross-Pitaevskii) equation \cite{Gross:1961} \cite%
{Pitaevskii:1961} ($\hbar $=1) 
\begin{equation}
i\frac{d\psi _{j}}{dt}=\frac{\partial H}{\partial \psi _{j}^{\ast }}
\label{twolevel2}
\end{equation}%
where $\psi _{j}=\psi _{j}(t)$ ($j=1,2$) are complex amplitudes of the wave
function $\psi $, and $H$ is the system energy given by 
\begin{equation}
H=\frac{V}{2}(\psi _{1}^{\ast }\psi _{2}+\psi _{1}\psi _{2}^{\ast })+\frac{R%
}{2}(\psi _{1}^{\ast }\psi _{1}-\psi _{2}\psi _{2}^{\ast })-\frac{c}{4}(\psi
_{1}^{\ast }\psi _{1}-\psi _{2}\psi _{2}^{\ast })^{2}  \label{twolevel1}
\end{equation}

Here $V$ is the strength of coupling between two modes, while $c$ is the
strength of interaction between atoms, $R$ characterizes the energy
difference between two wells (bias). \ For the purpose of present work, it
can be assumed that $R=0$ \cite{Giovanazzi:2008NJPhys}; as a result,
equation (\ref{twolevel2}) gives 
\begin{eqnarray}
i\frac{d\psi _1}{dt} &=&\frac V2\psi _2-\frac c2(\psi _1^{*}\psi _1-\psi
_2\psi _2^{*})\psi _1  \notag \\
i\frac{d\psi _2}{dt} &=&\frac V2\psi _1+\frac c2(\psi _1^{*}\psi _1-\psi
_2\psi _2^{*})\psi _2  \label{twolevel3}
\end{eqnarray}

In the literature, this model can be found in different physical contents
and somewhat different notations \cite{Milburn:1997PhysRev}, \cite%
{Smerzi:1997}, \cite{Raghavan:1999}, \cite{Wu:2000PhysRev}, \cite%
{Liu:2002PhysRev}. In general terms, it describes the tunneling effect in
two coupled Bose-Einstein condensates.

The following complexification of transformation (\ref{nb4_P}) is applied
now, 
\begin{eqnarray}
\psi _{1} &=&\sqrt{\frac{1}{2}K(1+P)}\exp (\delta i)  \notag \\
\psi _{2} &=&-\sqrt{\frac{1}{2}K(1-P)}\exp [(\delta +\Delta )i]
\label{psi1_2}
\end{eqnarray}%
where, the normalization condition for probability, $|\psi _{1}|^{2}$ $+$ $%
|\psi _{2}|^{2}=1$, takes the form $K=1$, and physical interpretation of the
energy partition index $P$ may depend upon the problem formulation.

Substituting (\ref{psi1_2}) in (\ref{twolevel3}), gives equation for the
fast phase $\delta $, 
\begin{equation}
\dot{\delta}=\frac{c}{2}P+\frac{V}{2}\sqrt{\frac{1-P}{1+P}\cos \Delta }
\label{Pdelta_quant}
\end{equation}
and the following classical Hamiltonian system for the canonically conjugate
variables $P$-$\Delta $, 
\begin{eqnarray}
\dot{P} &=&-V\sqrt{1-P^{2}}\sin \Delta \equiv -\frac{\partial H_{eff}}{%
\partial \Delta }  \notag \\
\dot{\Delta} &=&-cP+V\frac{P\cos \Delta }{\sqrt{1-P^{2}}}\equiv \frac{%
\partial H_{eff}}{\partial P}  \label{nb8_quant}
\end{eqnarray}
where $H_{eff}$ is so-called effective Josephson Hamiltonian 
\begin{equation}
H_{eff}=-\left( \frac{1}{2}cP^{2}+V\cos \Delta \sqrt{1-P^{2}}\right)
\label{Heff_quant}
\end{equation}

More often, however, the term Josephson Hamiltonian is used for a simplified
version of (\ref{Heff_quant}), associated with the classical rigid pendulum
as discussed at the end of Section \ref{Joseph}. Note that substitution of (%
\ref{psi1_2}) directly into the original Hamiltonian gives 
\begin{equation}
H_{eff}=2H  \label{Heff=2H}
\end{equation}

Although the numerical factor in this relationship can be eliminated by
re-scaling the time variable, nevertheless the presence of such factor moves
transformation (\ref{psi1_2}) out of the class of canonical transformations.
This problem can be fixed \cite{Liu:2003PhysRevLett}, however, at cost of
losing the traditional form Josephson's system.

Despite the fact that the original system is quantum by formulation, the
effective model (\ref{nb8_quant}) through (\ref{Heff_quant}) is usually
qualified as a classical Hamiltonian system since the variables $P$-$\Delta $
can be simultaneously determined provided that it is true for their initial
values. System (\ref{nb8_quant}) is equivalent to (\ref{nb8P}) under the
conditions $\varepsilon \Omega =V$ and $\kappa =c/V$. Substituting $%
P(t)=-\sin \theta (t)$ in (\ref{nb8_quant}) and (\ref{Heff_quant}), and
eliminating the phase $\Delta $, gives the effective conservative
oscillator, similar to that was introduced for the case of coupled Duffing
oscillators, 
\begin{equation}
\ddot{\theta}+\left( H_{eff}+\frac{c}{2}\right) ^{2}\frac{\tan \theta }{\cos
^{2}\theta }-\frac{1}{8}c^{2}\sin 2\theta =0  \label{nb13_quant}
\end{equation}%
where the number $H_{eff}$ must be calculated by fixing a trajectory on the
plane $P-\Delta $ or $\theta -\Delta $ before the oscillator is (\ref%
{nb13_quant}) is solved.

Introducing the new time scale $p=\left( H_{eff}+c/2\right) t$ and the
parameter $\mu =(1/2)\left( 1+2H_{eff}/c\right) ^{-2}$enables one of using
the solution in terms of elementary functions obtained in Section \ref%
{actionangle} for the case of coupled Duffing oscillators.

\subsection{Further analogies}

\label{analogies}

In terms of the macroscopic phase difference $\Phi =\phi _{r}-\phi _{l}$ ,
corresponding to the right (r) and left (l) wells, and the atom number
difference $k=(N_{l}-N_{r})/2$, Josephson type equations for the case
symmetric double-well potential are obtained from Gross-Pitaevskii theory 
\cite{Smerzi:1997}, \cite{Raghavan:1999}

\begin{eqnarray}
\dot{k} &=&-E_{J}\sqrt{1-4k^{2}/N^{2}}\sin \Phi  \notag \\
\dot{\Phi} &=&E_{C}k+\frac{4k/N^{2}}{\sqrt{1-4k^{2}/N^{2}}}E_{J}\cos \Phi
\label{BEC1}
\end{eqnarray}
where $N=N_{l}+N_{r}$ is the total number of atoms; $E_{J}$ and $E_{C}$ and
are called tunneling and charging energies, respectively, measured in units $%
\hbar $. The charging energy arises from interatomic interactions, the
tunneling energy determines the maximum current $I_{J}=$ $E_{J}$.

Introducing notations, 
\begin{eqnarray}
P &=&\frac{2k}{N}=\frac{N_{l}-N_{r}}{N_{l}+N_{r}}  \notag \\
\Delta &=&\pi -\Phi  \label{BEC2}
\end{eqnarray}
brings system (\ref{BEC1}) exactly to the Hamiltonian form (\ref{nb8P})-(\ref%
{Heff}), after the following parameter substitutions

\begin{eqnarray}
\varepsilon \Omega &=&\frac{2E_{J}}{N}  \notag \\
\kappa &=&\frac{N^{2}}{4}\frac{E_{C}}{E_{J}}  \label{analogy}
\end{eqnarray}

Moreover, re-scaling the time variable as $\varepsilon \Omega t=\bar{t}$ in (%
\ref{nb8P}) and $2(E_{J}/N)t=\bar{t}$ in (\ref{BEC1}), brings both systems
to the same single parameter form \cite{Smerzi:1997}, \cite{Raghavan:1999} 
\begin{eqnarray}
\frac{dP}{d\bar{t}} &=&-\sqrt{1-P^{2}}\sin \Phi  \notag \\
\frac{d\Phi }{d\bar{t}} &=&\kappa P+\frac{P}{\sqrt{1-P^{2}}}\cos \Phi
\label{smerzi}
\end{eqnarray}%
where the strength of nonlinearity of Duffing system (\ref{kappa}) $\kappa $
becomes equivalent to the parameter $\Lambda $ \cite{Smerzi:1997}, the
energy partition index between the Duffing oscillators, $P$, is equivalent
to the fractional relative population, and the phase shift between the
Duffing oscillators, $\Phi $, is the quantum phase difference between the
right and left components.

In order to use the results sections \ref{EPOscillator}, \ref{linearcase},
and \ref{actionangle}, the fractional relative population $P$ must be
expressed through the phase angle $\theta $. This leads to the integral

\begin{equation}
H_{\theta }\equiv \frac{1}{2}\dot{\theta}^{2}+\frac{1}{2}\left( H_{eff}+%
\frac{NE_{C}}{4}\right) ^{2}\tan ^{2}\theta +\left( \frac{NE_{C}}{8}\right)
^{2}\cos 2\theta  \label{H_theta_quant}
\end{equation}
where $H_{eff}$ is the effective Hamiltonian of system (\ref{BEC1})
calculated on a fixed trajectory.

Introducing the new time variable $p$ and the action-angle variables (\ref%
{nb25}) brings oscillator to its Hamiltonian form (\ref{nb26}), where 
\begin{eqnarray}
p &=&\left( H_{eff}+\frac{NE_{C}}{4}\right) t  \label{p mu} \\
\mu &=&\frac{1}{2}\left( 1+\frac{4H_{eff}}{NE_{C}}\right) ^{-2}  \notag
\end{eqnarray}

Therefore, analytical solution (\ref{nb25}) through (\ref{nb29}) obtained in
Section \ref{actionangle} for the angle $\theta $, gives 
\begin{equation}
P=\frac{N_{l}-N_{r}}{N_{l}+N_{r}}=-\sin \theta =-\frac{\sqrt{2I+I^{2}}}{1+I}%
\sin \phi  \label{z-solution}
\end{equation}
where the action-angle variables, $I$ and $\phi $, are given by (\ref{nb27})
and (\ref{nb29}).

\subsection{Second quantization}

As discussed at the end of Section \ref{Joseph}, system (\ref{BEC1}) becomes
equivalent to the classical pendulum under the conditions $|P|\ll 1$ and $%
\kappa >>1$ whose physical meaning is revealed by (\ref{analogy}). As
result, Hamiltonian of system (\ref{BEC1}) is reduced to that of the
classical pendulum 
\begin{eqnarray}
H_{J} &=&\frac{E_{C}}{2}k^{2}-E_{J}\cos \Phi  \notag \\
\dot{k} &=&-\frac{\partial H_{J}}{\partial \Phi }=-E_{J}\sin \Phi
\label{HJ_pend} \\
\dot{\Phi} &=&\frac{\partial H_{J}}{\partial k}=E_{C}k  \notag
\end{eqnarray}

Except some technical advantages, such simplification does not seam to be
crucial from the standpoint of classical dynamics since both systems (\ref%
{BEC1}) and (\ref{HJ_pend}) are integrable. However, in contrast to (\ref%
{BEC1}), the Hamiltonian of system (\ref{HJ_pend}) has the quadratic form
with respect to the `linear momentum' $k$, which is important for the idea
of second quantization \cite{Pitaevski:2003} 
\begin{equation}
\hat{H}_{J}=-\frac{E_{C}}{2}\frac{\partial ^{2}}{\partial \Phi ^{2}}%
-E_{J}\cos \Phi  \label{HJ_quant}
\end{equation}

Further discussion and references on the transition from (\ref{HJ_pend}) to (%
\ref{HJ_quant}) can be found in \cite{Krahn:2009}. Note that the effective
Hamiltonian of energy partition oscillator (\ref{H_theta_quant}) is
quadratic with respect to the `linear momentum' $\dot{\theta}$ so that no
further simplification would be required from the standpoint of second
quantization.

\section{CONCLUDING\ REMARKS}

This work points to interdisciplinary links between the resonance phenomena
in classical nonlinear dynamics and tunneling effects considered in the
macroscopic quantum dynamics. It is shown that the classical dynamics of 1:1
resonance interaction between two identical linearly coupled Duffing
oscillators is equivalent to the symmetric (non-biased) case of
`macroscopic' quantum dynamics of two weakly coupled Bose-Einstein
condensates. The analogy develops through the BJJ equations, however,
reduced to a single conservative EP oscillator imposing no additional
assumptions. The derived oscillator is solvable in quadratures, furthermore
it admits asymptotic solution in terms of elementary functions after
transition to the action-angle variables. The energy partition and coherency
indexes are introduced to provide a complete characterization of the system
dynamic states through the state variables of the EP oscillator. In
particular, nonlinear normal and local mode dynamics of the original system
associate with equilibrium points of the oscillator. Additional equilibrium
points - the local modes - may occur on high energy level as a result of the
symmetry breaking bifurcation. Further, it is shown that vibrations near
nonlinear local modes of the coupled Duffing's oscillators are equivalent to
the macroscopic quantum self-trapping (MQST) effect. As discussed earlier 
\cite{Raghavan:1999}, there are two types of MQST, such as running-phase
MQST and $\pi $-phase MQST. While BJJ captures both types of MQST, the SJJ
rigid pendulum approximation ignores the $\pi $-phase MQST. Finally, it is
noticed that, since the Hamiltonian of EP oscillator is always quadratic
with respect its linear momentum, the idea of second quantization can be
explored in terms of BJJ without usual transition to the pendulum
approximation. The quantum version of oscillator (\ref{H_theta_quant})
allows for asymptotic limits with square well and sine-wave potentials,
corresponding to the classic vibro-impact and harmonic oscillators,
respectively.

\section*{APPENDIX}

As mentioned in the text, substituting (\ref{nb4}) in (\ref{nb3}) and then
solving the set of equations with respect to the derivatives, gives 
\begin{eqnarray}
\dot{K} &=&\varepsilon K\Omega \cos \theta \sin (2\delta +\Delta )+\frac{%
2\varepsilon \sqrt{K}}{\Omega }  \notag \\
&&\times \left[ f_{1}\cos \left( \frac{\theta }{2}+\frac{\pi }{4}\right)
\sin \delta -f_{2}\cos \left( \frac{\theta }{2}-\frac{\pi }{4}\right) \sin
(\delta +\Delta )\right]  \notag \\
&&  \notag \\
\dot{\theta} &=&\varepsilon \Omega \lbrack \sin \Delta -\sin \theta \sin
(2\delta +\Delta )]-\frac{2\varepsilon }{\sqrt{K}\Omega }  \notag \\
&&\times \left[ f_{1}\cos \left( \frac{\theta }{2}-\frac{\pi }{4}\right)
\sin \delta +f_{2}\cos \left( \frac{\theta }{2}+\frac{\pi }{4}\right) \sin
(\delta +\Delta )\right]  \notag \\
&&  \TCItag{A1}  \label{A1} \\
\dot{\Delta} &=&-\frac{2\varepsilon }{\sqrt{K}\Omega }\left\{ \Omega ^{2}%
\sqrt{K}\cos \delta \cos (\delta +\Delta )\tan \theta \right.  \notag \\
&&+\left. \sec \theta \left[ f_{1}\cos \left( \frac{\theta }{2}-\frac{\pi }{4%
}\right) \cos \delta +f_{2}\cos \left( \frac{\theta }{2}+\frac{\pi }{4}%
\right) \cos (\delta +\Delta )\right] \right\}  \notag \\
&&  \notag \\
\dot{\delta} &=&\Omega \left[ 1+\varepsilon \cos \delta \cos (\delta +\Delta
)\tan \left( \frac{\theta }{2}+\frac{\pi }{4}\right) \right]  \notag \\
&&+\frac{\varepsilon }{\sqrt{K}\Omega }f_{1}\cos \delta \sec \left( \frac{%
\theta }{2}+\frac{\pi }{4}\right)  \notag
\end{eqnarray}
where $f_{i}=U^{\prime }(u_{i})$ and the coordinates $u_{i}$ ($i=1$,$2$) are
given by (\ref{nb4}).

System (\ref{A1}) is still an exact equivalent to system (\ref{nb3}) and
represents a standard dynamic system with a single fast phase, $\delta $.
Applying the standard averaging procedure over the fast phase $\delta $ on
the right-hand side of system (\ref{A1}), gives equations (\ref{nb8}).

\bibliographystyle{plain}
\bibliography{landauref}

\pagebreak

\includegraphics[scale=0.4]{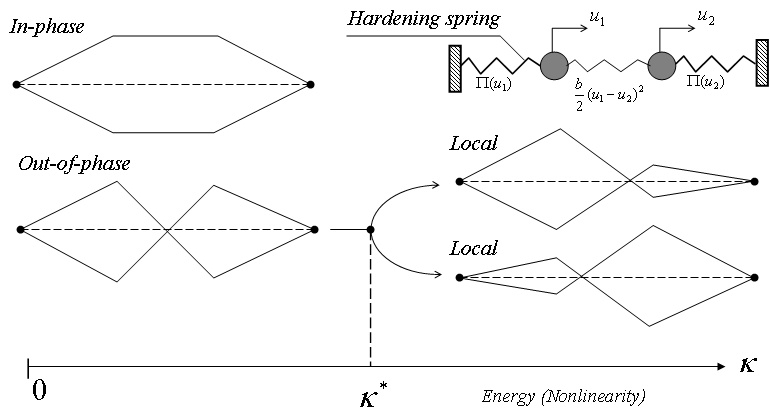} \smallskip

{\small FIG.1. Nonlinear normal and local modes of the coupled Duffing
oscillators with symmetry breaking bifurcation generating two stable local
modes from the out-of-phase mode.}

\bigskip

\includegraphics[scale=0.5]{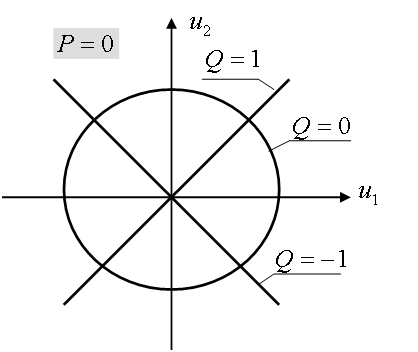} \smallskip

{\small FIG.2. Geometrical interpretation of the coherency index on
configuration plane in case of the energy equipartition, }$P=0${\small .}

{\small \bigskip }

\includegraphics[scale=0.75]{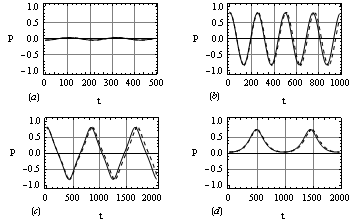} \smallskip

{\small FIG.3. Time histories of the energy partition index on the
supercritical nonlinearity (energy) level }$\kappa =1.2${\small \ at
different initial configurations of the model; see Fig. 4 for the
corresponding trajectories on configuration plane.}

{\small \bigskip }

\includegraphics[scale=0.75]{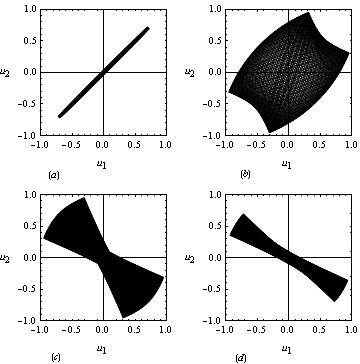} \smallskip

{\small FIG.4. Configuration plane trajectories on the supercritical energy
level, }$\kappa =1.2${\small , at different initial configurations of the
model: (a) and (b) - the dynamics in small and large neighborhoods of the
in-phase mode, respectively, (c) - the global dynamics around the unstable
out-phase and both stable local modes, and (d) - the neighborhood of one
local mode; the corresponding energy partition index is shown in Fig. 3 (a)
through (d), respectively, see also Fig. 5.}

{\small \bigskip }

\includegraphics[scale=0.75]{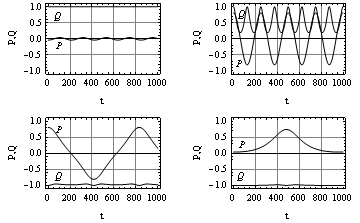} \smallskip

{\small FIG.5. Energy partition and coherency indexes on the supercritical
energy level, }$\kappa =1.2${\small , at different initial configurations of
the model; see Fig. 4 for the corresponding trajectories on configuration
plane.}

{\small \bigskip }

\includegraphics[scale=0.75]{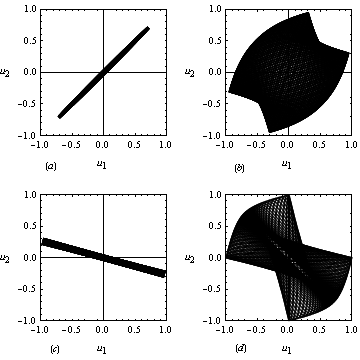} \smallskip

{\small FIG.6. Configuration plane trajectories on the supercritical energy
level, }$\kappa =2.0${\small : (a) and (b) - the dynamics in small and large
neighborhoods of the in-phase mode, respectively, (c) - the neighborhood of
local mode, and (d) - the drift over unstable out-of-phase mode and two
stable local modes; see Fig. 7 for the energy partition and coherency
indexes.}

\bigskip

\includegraphics[scale=0.75]{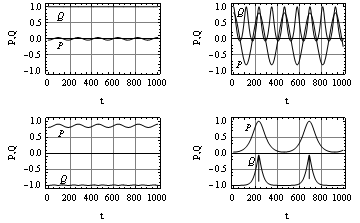} \smallskip

{\small FIG.7. Energy partition and coherency indexes on the supercritical
energy level, }$\kappa =2.0${\small , corresponding the dynamics illustrated
by Fig. 6.}

\bigskip

\includegraphics[scale=0.75]{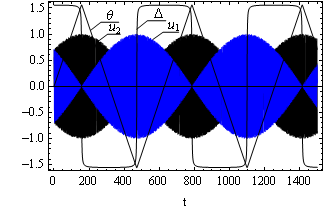} \smallskip

{\small FIG.8. Exact solutions for the beat dynamics of two identical
linearly coupled harmonic oscillators and associated phase variables of the
EP oscillator; highly intensive energy exchage.}

{\small \bigskip }

\includegraphics[scale=0.75]{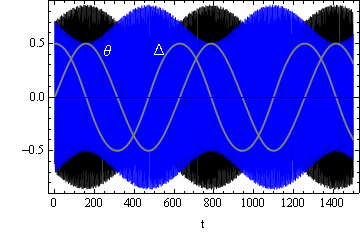} \smallskip

{\small FIG.9. Same as Fig. 8; moderate energy exchange.}

\bigskip

\includegraphics[scale=0.75]{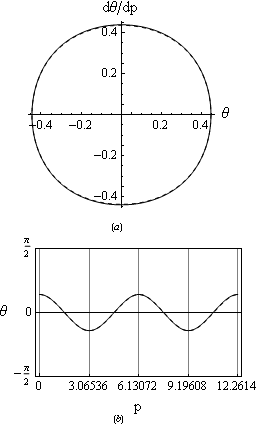} \smallskip

{\small FIG.10. Analytical and numerical solutions represented by solid and
dashed lines, respectively, for }$\kappa =0.591728${\small \ and }$%
P_{0}=-0.430443${\small ; (a) phase trajectory, and (b) time history; }$%
p=\varepsilon \Omega Gt${\small .}

\bigskip

\includegraphics[scale=0.75]{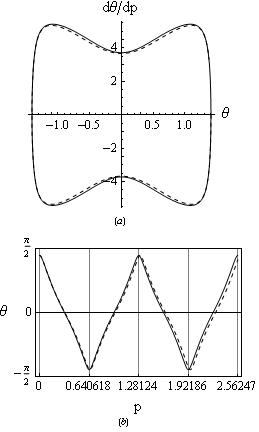} \smallskip

{\small FIG.11. Analytical and numerical solutions represented by solid and
dashed lines, respectively, for }$\kappa =1.41012${\small \ and }$%
P_{0}=-0.986701${\small ; (a) phase trajectory, and (b) time history; }$%
p=\varepsilon \Omega Gt${\small .}

{\small \bigskip }

\includegraphics[scale=0.75]{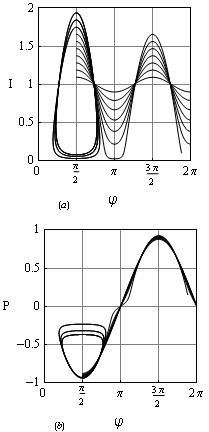} \smallskip

{\small FIG.12. Angle trapping of the action-angle variables at high action
levels (a), and its interpretation through the energy partition index (b); }$%
J=1.0${\small ; }$\mu =0.5${\small , }$1.0${\small ,...,}$5.0${\small .}

{\small \bigskip }

\includegraphics[scale=0.75]{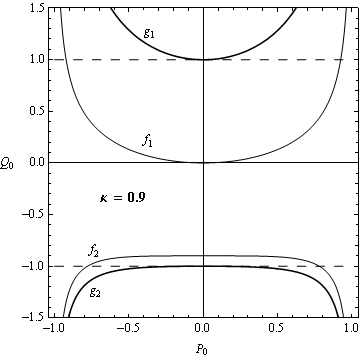} \smallskip

{\small FIG.13. The plane of initial coherence versus energy partition
showing no localization area at relatively low nonlinearity }$\kappa $%
{\small ; dashed horizontal lines bound the allowed region }$|Q_{0}|\leq 1$%
{\small .}

\bigskip

\includegraphics[scale=0.75]{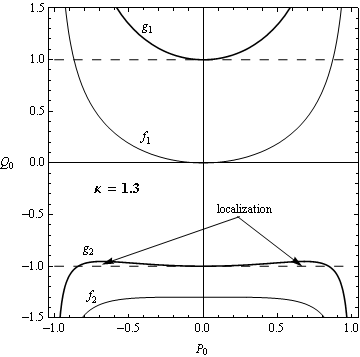} \smallskip

{\small FIG.14. The plane of initial coherence versus energy partition
showing two localization areas on higher nonlinearity level }$\kappa $%
{\small .}

\bigskip

\includegraphics[scale=0.5]{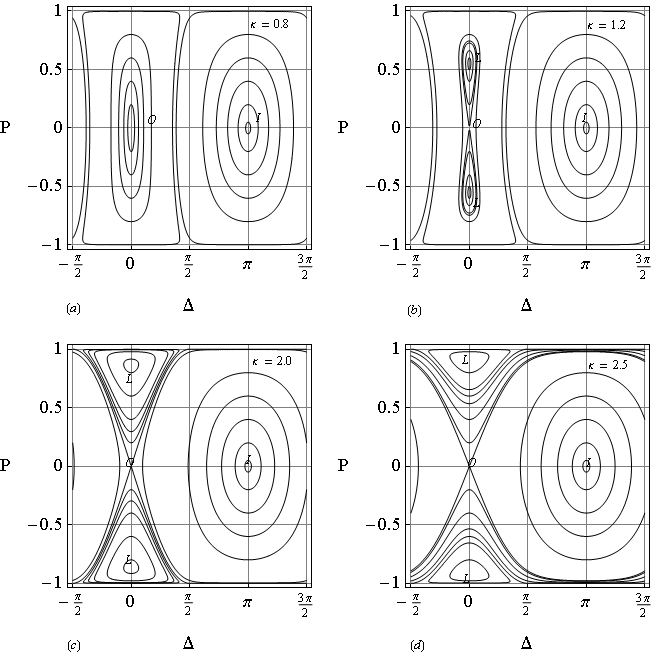} \smallskip

{\small FIG.15. Contour lines of the effective Hamiltonian of the energy
exchange equations between Duffing oscillators at different nonlinearity
levels: (a) quasi linear case, (b) post-bifurcation case }$\kappa =1.2$%
{\small \ with center-saddle bifurcation near the out-of-phase mode, (c)
strongly nonlinear multiple (\TEXTsymbol{>}2) modes effect, and (d) strongly
nonlinear effect of separatrix topological transition.}

{\small \bigskip }

\includegraphics[scale=0.5]{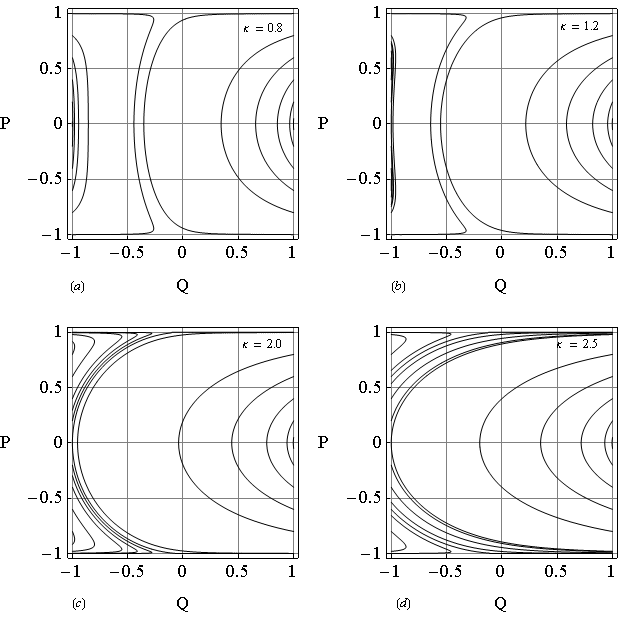} \smallskip

{\small FIG.16. Energy partition versus coherency index diagrams at
different nonlinearity levels: (a) quasi linear case, (b) post-bifurcation
case }$\kappa =1.2${\small \ with center-saddle bifurcation near the
out-of-phase mode, (c) strongly nonlinear multiple (\TEXTsymbol{>}2) modes
effect, and (d) strongly nonlinear effect of separatrix topological
transition.}

\end{document}